# De doos van Pandora/La boîte de Pandora

*Rubriek gewijd aan archiefvondsten, instrumentbeschrijvingen, e.d./*
*Rubrique consacrée aux trouvailles d'archives, aux descriptions d'instruments, etc.*

## Investigations into the origin of Einstein's Sink

ALEXANDER G.M. PIETROW*

ABSTRACT

Einstein's sink is a well-known object among physics and astronomy students at Leiden University. Stories about its origin have been passed down since it was moved to the large lecture room of the then newly built Oort Building in 1998. These stories claim that it was seen as an Einstein relic by the physics faculty, kept close to inspire young minds. After researching this story, I found that the sink is from the early 20th century and that it once stood in the large lecture room of the Kamerlingh Onnes Laboratory in Leiden. It is likely that at this time it was used by many scientists, including Einstein. However, my research also shows that the move from the KOL was done purely for financial reasons and had nothing to do with Einstein. It is possible that the story was made up to conceal this fact, creating a more appealing reason for these cutbacks. The sink is seen by many as a connection to Einstein's and the university's past and is liked for this fact and not the aforementioned legend.

## Introduction

Leiden has a rich history of scientific discoveries in almost every field and has been a hotspot for many famous scientists ever since its university was established in 1575.[1] The presence of famous scientists, combined with a significant number of historical objects that are littered across the university's science faculty buildings, has led to a strong oral tradition among students. Due to the transient nature of a student's educational career, rarely lasting more than half a decade, old stories tend to evolve rapidly and the facts on which they are based become obscured. However, these stories are usually based on a true story that can be uncovered with some research, providing the opportunity to gain insight into the formation of academic myths. In this article, we wish to focus on one such story, that of Einstein's Sink.

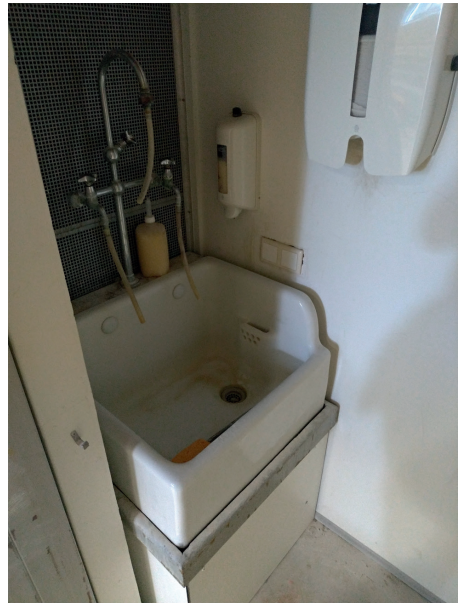

Fig. 1: Einstein's Sink at its current location in the De Sitter room.

* Alexander G.M. Pietrow, Department of Astronomy, Stockholm University. E-mail: alex.pietrow@astro.su.se. The Institute for Solar Physics is supported by a grant for research infrastructures of national importance from the Swedish Research Council (registration number 2017-00625).

When I started out as a physics and astronomy undergraduate student in 2010, it did not take long until I was informed about the whereabouts of Heike Kamerlingh Onnes's electric kettle – displayed in front of the De Sitterzaal, next to one of his helium liquefiers. I was also told that the telescope that stood near the entrance of the Huygens building was once the biggest in the world, that Ejnar Hertzsprung named his daughter Rigel (after the star), and of course that the main lecture room sink was once used by Albert Einstein.[2] More specifically, the legend was that Einstein once washed his hands in it after a colloquium in the old physics laboratory, renamed the Kamerlingh Onnes laboratory (KOL) in 1932. From that moment on the sink was seen as an essential part of the lecture room, to the extent that it was transferred to the new Huygens Laboratory during the move in 1998. Some versions of the story even added that the sink was moved to inspire new students by cleaning the blackboards with the sponge water of Einstein.

Conversations with students from various cohorts revealed that many versions of the story circulated simultaneously, suggesting that the story spreads across a generation of students and then trickles down to the new freshmen a year after where it spreads once more. Some say that Einstein washed his hands in the sink only once, others that he did it many times. Some stories say that the move of the sink was requested by the director, while others say that it was done secretly as a student prank or by a staff member who was an Einstein admirer. Of course, none of these stories were referenced, and everyone said that they had heard it from a fellow student or a professor. However, people did seem to believe that part of the story was true, as Leiden possesses several other 'Einstein relics', such as Einstein's Pen in Museum Boerhaave, Einstein's chair in the Old Observatory and the House of Ehrenfest, famous for Einstein living there. After hearing this story several times from different people, it was possible to extract two elements that were always present, namely that the sink was used by Einstein and that the sink was moved to the Huygens Laboratory because of this fact.

It is interesting to see a seemingly mundane object gaining such long-lasting fame and status. It does not serve any purpose to the university administrators, students or professors, other than that of providing water to clean the blackboard and wash one's hands. There are no rituals based around it, nor does the university itself consider it an important relic from its past. However, the story of the sink somehow still manages to capture the imagination of students and staff members alike, spreading to new generations each year and staying mostly the same for nearly half a century.

Scientific 'relics' are not an uncommon sight in academia and seem to often be possessions of famous scientists, usually having nothing to do with their work or achievements, but rather come from their everyday lives. These objects usually lost their usefulness and are kept purely because of their connections to past users. Leiden is full of these kind of relics, such as the aforementioned electric kettle, chair and pen but furthermore there is the desk of Ejnar Hertzsprung and Jan Oort's wooden bench, which are used in Leiden Observatory, Henk van de Hulst's tap and even the houses of Ehrenfest and Frederik Kaiser. Abroad we find similar objects like Sir Isaac Newton's house in Lincolnshire, Albert Einstein's blackboard and chalk in Nottingham, Edwin Hubble's basketball in Chicago, Carl Gauss' cap in Gottingen and even the potato masher of Ernest Rutherford in London.[3] In addition, there are the morbid middle finger of Galileo in Florence, the dying breath of Edison in Michigan and much more.[4] Keeping and displaying such items reminds us of the Catholic Church's practice of relic collecting, which seems to be at odds with the objectivity and distance that scientists strive for in their work and image. Therefore, I would also like to investigate why items like these are so popular amongst scientists and students alike.

I tried to find out whether there is any truth behind the legend of Einstein's sink and looked into why exactly it is this object that became so famous by answering the following questions. First, is the sink old enough to have been used by Einstein? Second, has it ever been at the KOL? Was it at a location where Einstein could have used it? And finally, why is it so important to the faculty?

I will answer these questions in this article and draw a conclusion based on the available information. Furthermore, we will cover the more recent history of the sink and the petition to have it moved to the new science campus.

*Origin of the Sink*
After comparing the sink to other models found in the Huygens Laboratory and even the much older Gorlaeus Laboratory, which was built in the 1960s, I was not able to find anything that looked like this sink. Moreover, some numbers were visible on the bottom of the sink, one of them suggesting that it stems from 1923. (fig. 2) After having confirmed that this sink model is indeed unique at the campus, I decided to ask for the opinion of several antique dealers. Most of the replies were positive, confirming that the sink was indeed from the early 20th century. Most notable was the stock at a shop for old building materials called 'LEEN' that had several sinks with similar soap hole patterns to that of our sink.[5] According to the owner, these sinks were all from the early 20th century and made either by the Dutch company Sphinx or imported from the United Kingdom.[6] Beyond that it proved difficult to find the exact origin of the sink. Sphinx did not have a catalogue going back this far and the dedicated earthwork association 'Vereniging Maastrichts Aardewerk' could not give a confirmation either way. Other candidates were the German 'Viega', and the British 'Burlington'. Both denied ever making such products. The owner of LEEN did recognize our sink as being from the same era, confirming the possibility that it is old enough to have been used by Einstein.

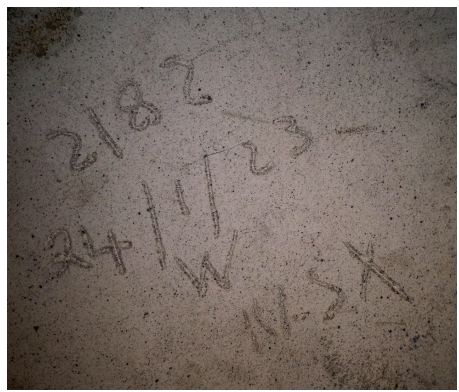

Fig. 2: A picture of the underside of the sink, showing some kind of serial number and possibly a date. '24/1/23'.

*The sink in the KOL*
The Kamerlingh Onnes Laboratory was built in 1856 as a physical, chemical and anatomical laboratory based on the design of the Royal architect H.F.G.N. Camp. It was built in the center of Leiden on the clearing that was created during the Leiden gunpowder disaster of 1807. This was to the great discontent of locals, who were afraid of another explosion in this strange lab. In 1932 the building was named after Heike Kamerlingh Onnes (1853–1926), who won the Nobel Prize in 1913 for liquefying helium. Because of this, the building was often nicknamed the 'coldest place on earth.' Over the years the KOL was expanded to accommodate the growing number of scientists within until they were all moved out in 1998.

It is fortunate that the building and its inhabitants were so influential, because this led people to take photographs in a time when this was still done sparingly. These can be found at the Leidse beeldbank (the city's photographic archive), the University and Rijksmuseum Boerhaave. With this information, it proved remarkably easy to confirm whether or not the sink had ever been in the KOL because it could be seen in several photographs at this location. We find the sink in the large lecture room which was built in 1922 and see it popping up in several photos taken at later times.[7] The sink seems to be part of a mirrored set that stood at the end of a long row of experimental tables. Other pictures from this archive show no similar sinks in the building. We believe that this is because of the modular expansion of the building, proving that this is the only place in the KOL where this type of sink can be found.

From these pictures, we can conclude that the sink on the right-hand side of the lab is, in fact, what has become to be known as Einstein's sink and that it had to be from at least 1922, which matches the predictions of the antique dealers. The discrepancy of one year between the date on the sink and the date on the photo can be explained with the fact that the dates of these photographs were added after the war and could easily be off by a year or two.[8] Nevertheless, it places the sink in the right time.

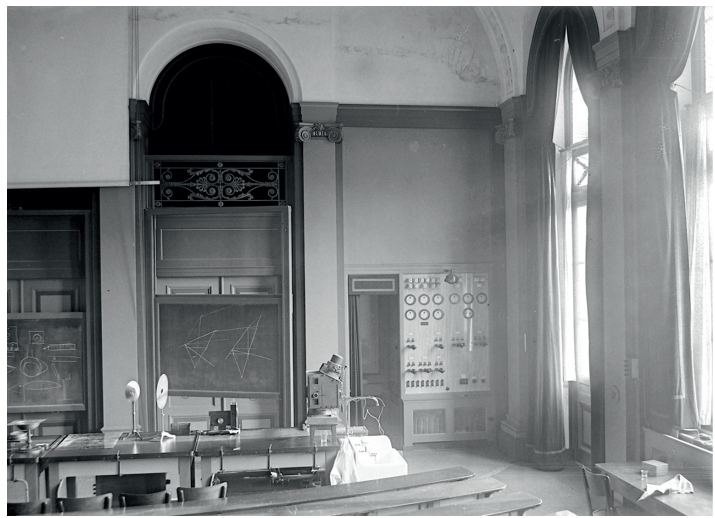

Fig. 3: A picture of the right side of the large lecture room of the KOL, taken shortly after the completion of the room in 1922. The sink is clearly visible on the lower central part of the picture, to the right of the tables. Photo: Kamerlingh Onnes Laboratory.

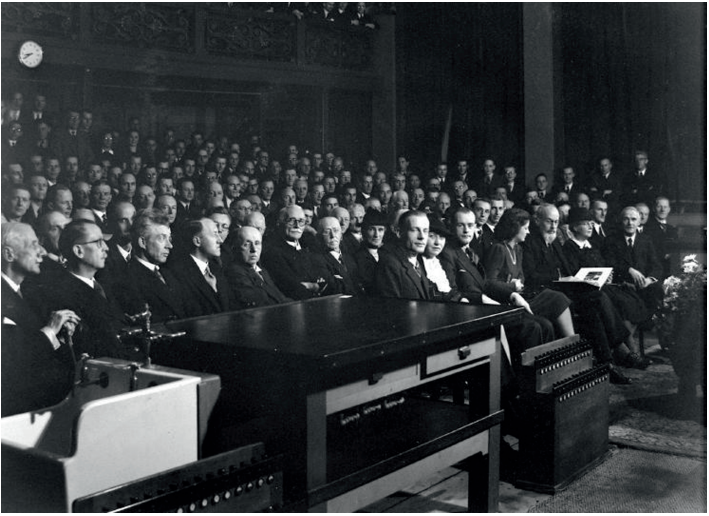

Fig. 4: A photograph taken during the retirement ceremony of Prof. Crommelin in 1944. The sink is clearly visible in the lower left corner. Photo: Kamerlingh Onnes Laboratory.

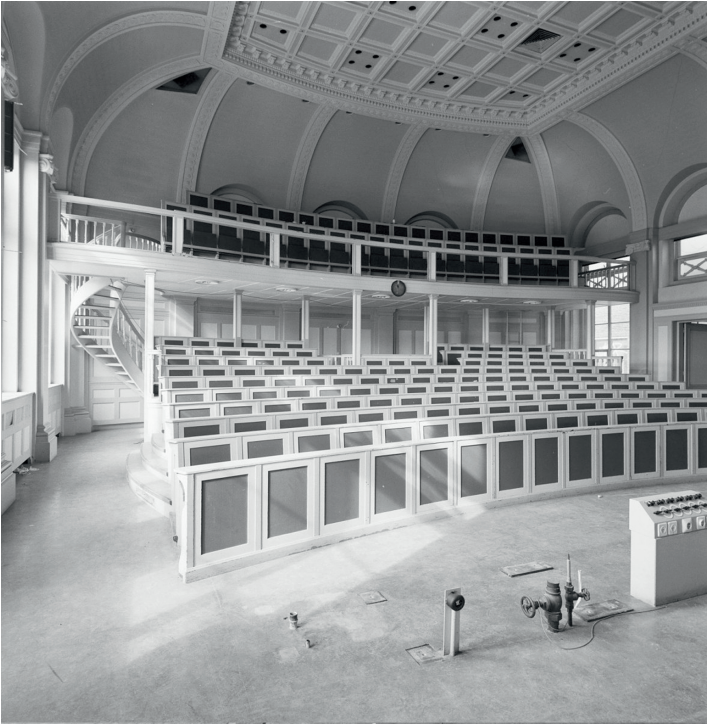

Fig. 5: A photo of the large lecture room taken in 2002, showing that the sink had indeed been removed by then. Photo: Rijksdienst voor het Cultureel Erfgoed.

*Investigations into the origin of Einstein's Sink*

*Einstein at the KOL*

Einstein had very close friendships with several Leiden researchers, primarily Paul Ehrenfest, but also Heike Kamerlingh Onnes and Hendrik Antoon Lorentz. He was said to enjoy Leiden because of the relaxed fashion in which he could discuss fundamental physics there. This presented a contrast to the rigid atmosphere that was present at his department in Berlin, where he worked and lived from 1914 to 1932.[9] Because of this Einstein would come to Leiden regularly to speak with his colleagues and to teach.

In October 1920, he became a professor by special appointment (*bijzonder hoogleraar*); this allowed Einstein to have what he described as a 'comet-like existence in Leiden,' where he would regularly visit for short periods of time. Einstein's last visit to Leiden took place in April 1930, shortly before his permanent move to the United States. He kept his title until 1946, but never visited after he moved to the US.

The large lecture room was built shortly after Einstein's appointment as professor; it was located in the same building where his close friends and colleagues worked. On top of that, it is known that Einstein gave lectures to both students and physicists during his time there. We therefore think that despite there being no written or photographic evidence of Einstein's ever entering this room, it is unlikely that he did not. Given his fame and popularity in the 20s, he would have drawn big crowds whenever he would give a talk; they had to be seated somewhere. On top of that, it is highly likely to anyone who has ever used chalk that he must have washed his hands after the lectures. Based on the assumption that he would fill the blackboards from left to right and go for the closest sink to wash his hands, it is most likely that the right-hand sink would have been used for that. Therefore, it is very probable that Albert Einstein washed his hands in the sink at least once during his many visits to Leiden. On top of that, for the same reason, we can assume that the sink was also used by Kamerlingh Onnes, Lorentz, Ehrenfest and many visiting physicists.

*Moving the Sink*

In 1974 a part of the physics department moved to the newly built Huygens Laboratory in the Bio Science Park, splitting the department in two. After a few years, it was decided that this split was not productive for science, and in 1985 plans were made to bring the rest of the physics department to the new location. To achieve this a new building was commissioned and built right next to the Huygens Building. When this building was completed in 1998, there was finally room for the entire physics department, and the KOL could be abandoned altogether. However, because of financial difficulties, as much equipment and furniture as possible was brought over from the old laboratory.[10] Thus, financial motives were the only ones that played a role in the move of the sink. As further confirmation, the sink was not the only thing taken from the KOL and brought to the De Sitterzaal. The demonstration tables and blackboards were brought over for the same reason. Thus, the table and blackboards were also used by Einstein, but for some reason, they did not get associated with him.

After the physicists moved out of the KOL, it fell into neglect and decay until it was renovated in 2004 and taken into use by the law faculty. Most of the old furniture was replaced, including all the old sinks.[11]

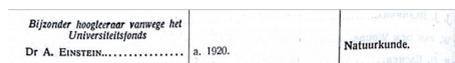

Fig. 6: Excerpt from 'Gids der Rijksuniversiteit te Leiden voor 1939/'40' listing Einstein as a professor by special appointment. Photo: J. Wasala.

The second sink was probably lost during these renovations, which means that our sink, blackboards and experimental tables are possibly the last remaining items from the KOL that have been used directly by the famous scientists who worked there.

*Why do people care about the sink?*
Despite his own wishes, Einstein reached an almost god-like status. He is known by nearly everyone and is generally remembered as an exceptionally brilliant, larger than life personality, to such an extent that some argue that these 'super models of science' raise the bar for people to enter science.[12] Objects like Einstein's sink, Kamerlingh Onnes' electric kettle or Gauss' cap can be seen as relics that support this hero worship. However, they can also be seen as items that are kept to humanize these heroes and show that they were not above mundane problems like having dirty hands, needing warm water or having a cold head. The sink itself never played a crucial role in Einstein's work or that of the physics department in general. It was not kept after the move because of its relation to Einstein, but rather because money could be saved. Yet this is not how the sink is remembered and instead it somehow managed to become an academic heirloom. Just like personal possessions and family heirlooms, these objects create a sense of community and connection to the past.[13] On top of that the story ads a certain romanticism to the move to the new building by obscuring the cutbacks. A department that can't afford a sink is generally not in a good place, which is not something that generally gets admitted. A way around this is finding a more appealing reason to reuse the old sink.

*The Sink now and in the future*
A new science campus that would house all scientists from the nearby buildings was commissioned in 2008. Talk about the new campus was increasing when in late 2015 plans were being made for moving labs and equipment. To the dismay of some students, the sink was not amongst these items. In response, a petition to move the sink to the new campus was started by the historical committee of the astronomical study association Leidsch Astronomisch Dispuut 'F. Kaiser'.[14] The petition caught the attention of local media like university newspaper *Mare* and newspaper *Leidsch Dagblad*, but also of two national magazines, the *Nederlands Tijdschrift voor Natuurkunde* (Dutch physics journal) and the *Historisch Nieuwsblad* (historical news journal).[15] It was also mentioned during Rijksmuseum Boerhaave's 'Einstein bicycle tour'.[16] This attention, along with some pushing on social media resulted in nearly 200 people signing the petition. The comments that were posted alongside the signatures varied in length and seriousness. However, the most recurring theme was that of the sink being a connection to Einstein and his past in Leiden.

The request to move the sink was submitted to the faculty board on April 1st, 2015. On the 21st of April, the board replied positively to the request, promising to move the sink to the new building once the plans for the second phase were complete. In the meantime, a small plaque was added to the sink. During the petition drive, the sink obtained more of a cult status among the students, and it appeared on several web pages like Wikipedia (Currently in 8 languages), TripAdvisor, where it is currently #65 on the list of things to do in Leiden, and several other sites. Tourists have come for tours/visits and there even exists 'Sink merchandise'.

As of December 2018, no additional plans have been made by the faculty board because the new campus building got delayed. If no further delays come up, we can expect plans on internal decorations to be made in 2021.

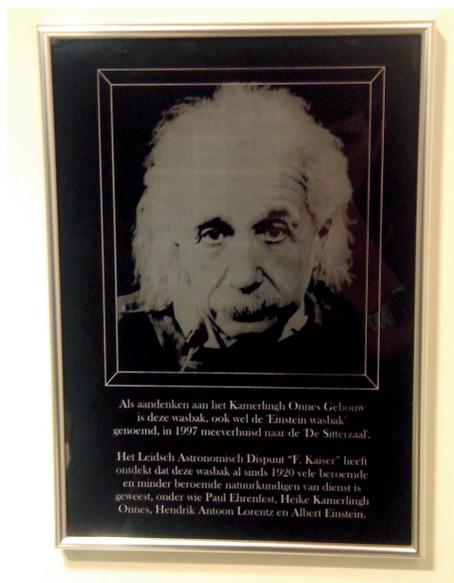

Fig. 7: The sign that was put up after the success of the petition to save Einstein's Sink. Loosely translated it reads: 'This sink, also known as 'Einstein's Sink' was taken from the Kamerlingh Onnes building to the De Sitter Room in 1997 as a memento to the old building. The Leids Astronomisch Dispuut 'F. Kaiser' has discovered that many famous and less famous scientists washed their hands in this sink, amongst them Paul Ehrenfest, Heike Kamerlingh Onnes, Hendrik Antoon Lorentz and Albert Einstein.'

*Conclusion*
During my research, I found out that the sink is indeed old enough to have been used by Einstein, and that it has stood in the large lecture room of the KOL from 1923 to 1998, and so had seven years of overlap with Einstein's time in Leiden. I also concluded that it is very likely that Einstein used the sink because he gave lectures and the largest lecture room would be a logical location for them, given his popularity. After its extended stay at the KOL, the sink was indeed moved from its original location to the De Sitter room in the Oort building. However, this had nothing to do with Einstein or anyone else using it, but was done only because it was cheaper than buying a new sink. It is possible that the link between the sink and Einstein was made up to conceal this very fact, but it could also be a complete coincidence. Nevertheless, the sink is a popular among students and staff alike and even garnered some attention outside of the university due to its connection to Einstein. Like other scientific heirlooms across the world, I believe that the sink humanizes its users and lowers the point of entry for people to do science. This connection to Einstein has played a significant role in the petition and is the reason why it will be moved to the new campus in the future. This means that the student legend that was passed down for so long was only partially correct. However, the story will become more accurate after the future move, which will be done solely for its connection to Einstein.

*Acknowledgments*
I would like to thank Prof. Frans van Lunteren, Prof. Dirk van Delft and Anna-Luna Post for pointing me to relevant archival photos and for their comments, which significantly improved the manuscript. Furthermore, I wish to thank Prof. P. Kes, Prof. R. de Bruyn Ouboter, Prof. J.H. van der Waals and Drs. T.M. van Beek for taking the time to answer my questions concerning this research. Finally, I would also like to extend my gratitude to the Leidsch Astronomisch dispuut 'F. Kaiser' for hosting the petition website.

*Noten*
1 *Van kabinet naar science park: 200 jaar Faculteit der Wiskunde en Natuurwetenschappen* (Leiden 2015).
2 Huib J. Zuidervaart, '"Zo'n mooie machine, waarvan de kwaliteit door alle astronomen wordt erkend": Een biografie van een vrijwel niet gebruikte telescoop', *Gewina: Tijdschrift voor de Geschiedenis der Geneeskunde, Natuurwetenschappen, Wiskunde en*